\documentclass[12pt,english,a4paper]{article}
\usepackage[T1]{fontenc}
\usepackage[latin9]{inputenc}
\usepackage{verbatim}
\usepackage{amsmath}
\usepackage{graphicx}

\makeatletter

\usepackage{float}

\usepackage{epsfig}

\makeatother

\usepackage{babel}

\begin{document}

\makeatother

\title{C\textbf{\large orrections to gauge theories in effective quantum
gravity with a cutoff}}

\author{G. Cynolter and E. Lendvai}

\date{MTA-ELTE Research Group in Theoretical Physics, Eötvös University,
Budapest, 1117 Pázmány Péter sétány 1/A, Hungary }
\maketitle
\begin{abstract}
We calculate the lowest order quantum gravity contributions to QED
beta function in an effective field theory picture with a momentum
cutoff. We use a recently proposed 4 dimensional improved momentum
cutoff that preserves gauge and Lorentz symmetries. We find that there
is non-vanishing quadratic contribution to the photon 2-point function
but that does not lead to the running of the original coupling after
renormalization. We argue that gravity cannot turn gauge theories
asymptotically free.
\end{abstract}

\section{Introduction}

The LHC experiments discovered a bosonic resonance consistent with
the SM (Standard Model) Higgs boson with a mass approximately 126
GeV \cite{atlas,cms}. This Higgs mass implies that the SM is renormalizable
and might be valid up to the scale of the Planck mass \cite{stabil,misa}
and we live in a metastable world \cite{stab3}. Only few evidences
demand physics beyond the SM. The origin and properties of possible
new physics motivated by dark matter and observed baryon asymmetry
of the universe are still unknown. Considering the SM or its extensions
valid up to the Planck scale gravitational corrections are present
and can be estimated. As Einstein's general relativity includes a
dimensionful constant $\kappa$ with negative mass dimension ($\kappa^{2}=32\pi G_{\mathrm{N}}=1/M_{P}^{2}$),
its perturbative quantization leads to a non-renormalizable theory,
implying that the cutoff of the theory cannot be taken to infinity.
Donoghue argued that assuming there is some yet unknown, well defined
theory of quantum gravity that yields the observed general relativity
as a low energy limit, the Einstein-Hilbert action can be used to
calculate gravitational correction in the framework of effective field
theories well below the Planck mass \cite{greff,greff2}.

The effective field theory treatment was recently used to study quantum
corrections to gauge theories. Robinson and Wilczek claimed that quadratically
divergent contribution to the Yang-Mills beta function is negative
and points toward asymptotic freedom \cite{robinson}. There were
several controversial results about this statement in the literature.
Pietrykowski showed in \cite{pietr} that in the Maxwell-Einstein
theory the result is gauge dependent. Toms repeated the calculation
in the gauge choice independent background field method using dimensional
regularization and has found no quadratically divergent gravity contribution
to the beta function \cite{toms1}. Diagrammatic calculation employing
dimensional regularization and naive momentum cutoff \cite{rodigast}
found vanishing quadratic contribution and the logarithmic divergences
renormalize dimension-6 operators in agreement with the early results
of Deser et al. \cite{deser}. Toms later applied proper time cutoff
regularization and claimed that quadratic dependence on the energy
remains in the QED one-loop effective action \cite{toms2}. Analysis
using the background field method employing the Vilkowsky-De Witt
formalism \cite{wu,pietr2} and special loop regularization that respects
Ward identities both found non-vanishing quadratic contributions to
the beta function \cite{wu} but with sign opposite to \cite{robinson,toms2}.
There are many various results, sometimes contradicting to each other
and the physical reality of quadratic corrections to the gauge coupling
was questioned \cite{ellis,don1,brazil}. The situation could be clarified
using a cutoff calculation respecting the symmetries of the models
and correctly interpreting the divergences appearing in the calculations.

Recently we developed a new improved momentum cutoff regularization
which by construction respects the gauge and Lorentz symmetries of
gauge theories at one loop level \cite{uj}. In this paper we apply
it to the effective Maxwell-Einstein system to estimate the regularized
gravitational corrections to the photon two point functions in the
simplest possible model.

The paper is organized as follows. In section 2. the improved momentum
cutoff is summarized, in section 3. the effective gravity contribution
to quantum electrodynamics is calculated, then the renormalization
is discussed. The paper is closed with conclusions and an appendix.

\section{Improved momentum cutoff}

A novel regularization of gauge theories is proposed in \cite{uj}
based on 4 dimensional momentum cutoff to evaluate 1-loop divergent
integrals. The idea was to construct a cutoff regularization which
does not brake gauge symmetries and the necessary shift of the loop-momentum
is allowed as no surface terms are generated. The loop calculation
starts with Wick rotation, Feynman-parametrization and loop-momentum
shift. Only the treatment of free Lorentz indices should be changed
compared to the naive cutoff calculation.

We start with the observation that the contraction with $\eta_{\mu\nu}$
(tracing) does not necessarily commute with loop-integration in divergent
cases. Therefore the substitution of \begin{equation}
k_{\mu}k_{\nu}\rightarrow\frac{1}{4}\eta_{\mu\nu}k^{2}\label{eq:negyed}\end{equation}
is not valid under divergent integrals, where $k$ is the loop-momentum%
{}. The usual factor $1/4$ is the result of tracing both sides under
a loop integral, e.g. changing the order of tracing and the integration.
In the new approach the integrals with free Lorentz indices are defined
using physical consistency conditions, such as gauge invariance or
freedom of momentum routing. Based on the diagrammatical proof of
gauge invariance it can be shown that the two conditions are related
and both are in connection with the requirement of vanishing surface
terms. It was proposed in \cite{uj} that instead of \eqref{eq:negyed}
the general identification of the cutoff regulated integrals in gauge
theories \begin{equation}
\int_{\Lambda\: reg}d^{4}l_{E}\frac{l_{E\mu}l_{E\nu}}{\left(l_{E}^{2}+m^{2}\right)^{n+1}}=\frac{1}{2n}\eta_{\mu\nu}\int_{\Lambda\: reg}d^{4}l_{E}\frac{1}{\left(l_{E}^{2}+m^{2}\right)^{n}},\ \ \ \ \ n=1,2,...\label{eq:idn}\end{equation}
will satisfy the Ward-Takahashi identities and gauge invariance at
1-loop ($l{}_{E}$ is the shifted Euclidean loop-momentum). In case
of divergent integrals it differs from \eqref{eq:negyed}, for non-divergent
cases both substitutions give the same results at ${\cal O}(1/\Lambda^{2})$
(the difference is a vanishing surface term). It is shown in \cite{uj}
that this definition is robust in gauge theories, differently organized
calculations of the 1-loop functions agree with each other using \eqref{eq:idn}
and disagree using \eqref{eq:negyed}. For four free indices the gauge
invariance dictates ($n=2,3,...$)\begin{equation}
\int_{\Lambda\: reg}d^{4}l_{E}\frac{l_{E\alpha}l_{E\beta}l_{E\mu}l_{E\rho}}{\left(l_{E}^{2}+m^{2}\right)^{n+1}}=\frac{1}{4n(n-1)}\int_{\Lambda\: reg}d^{4}l_{E}\frac{\eta_{\alpha\beta}\eta_{\mu\rho}+\eta_{\alpha\mu}\eta_{\beta\rho}+\eta_{\alpha\rho}\eta_{\beta\mu}}{\left(l_{E}^{2}+m^{2}\right)^{n-1}}.\label{eq:idn4}\end{equation}
For 6 and more free indices appropriate rules can be derived (or \eqref{eq:idn}
can be used recursively for each allowed pair). Finally the scalar
integrals are evaluated with a simple Euclidean momentum cutoff. The
method was successfully applied to an effective model to estimate
oblique corrections \cite{fcmlambda}.

There are similar attempts to define a regularization that respects
the original gauge and Lorentz symmetries of the Lagrangian but work
in four spacetime dimensions usually with a cutoff \cite{gu,wu1}.
Some methods can separate the divergences of the theories and does
not rely on a physical cutoff \cite{nemes,rosten,polon} or even could
be independent of it \cite{pittau}. For further literature see references
in \cite{uj}.

Under this modified cutoff regularization the terms with numerators
proportional to the loop momentum are all defined by the possible
tensor structures. Odd number of $l_{E}$'s give zero as usual, but
the integral of even number of $l_{E}$ are defined by \eqref{eq:idn},
\eqref{eq:idn4} and similarly for more indices, this guarantees that
the symmetries are not violated. The calculation is performed in 4
dimensions, the finite terms are equivalent with the results of dimensional
regularization. The method identifies quadratic divergences while
gauge and Lorentz symmetries are respected. We stress that the method
treats differently momenta with free ($k_{\mu}k_{\nu}$) and contracted
Lorentz indices ($k^{2}$), the order of tracing and performing the
regulated integral cannot be changed similarly to dimensional regularization.
The famous triangle anomaly can be unambiguously defined and presented
in \cite{tranom}.

However even using dimensional regularization one is able to define
cutoff results in agreement with the present method. In dimensional
regularization singularities are identified as $1/\epsilon$ poles,
power counting shows that these are the logarithmic divergences of
the theory. Naively quadratic divergences are set to zero in the process,
but already Veltman noticed that these divergences can be identified
by calculating the poles in $d=2$. Careful calculation of the Veltman-Passarino
1-loop functions in dimensional regularization and with 4-momentum
cutoff leads to the following identifications \cite{uj,hagiwara,harada}
\begin{eqnarray}
4\pi\mu^{2}\left(\frac{1}{\epsilon-1}+1\right) & = & \Lambda^{2},\label{eq: quad}\\
\frac{1}{\epsilon}-\gamma_{E}+\ln\left(4\pi\mu^{2}\right)+1 & = & \ln\Lambda^{2}.\label{eq:log}\end{eqnarray}
The finite terms are unambiguously defined \begin{equation}
f_{{\rm finite}}=\lim_{\epsilon\rightarrow0}\left[f(\epsilon)-R(0)\left(\frac{1}{\epsilon}-\gamma_{E}+\ln4\pi+1\right)-R(1)\left(\frac{1}{\epsilon-1}+1\right)\right],\label{eq:finite}\end{equation}
where $R(0)$, \, $R(1)$ are the residues of the poles at $\epsilon=0,\,1$
respectively. Using \eqref{eq: quad}, \eqref{eq:log} and \eqref{eq:finite}
at 1-loop the results of the improved cutoff can be reproduced using
dimensional regularization without any ambiguous subtraction.

This novel momentum cutoff regularization was developed and used in
gauge theories at one loop level and here it is applied to a simple
system where quadratically divergent gravitational corrections could
play a role.

\section{Effective Maxwell-Einstein theory}

The quantum corrections to the photon self energy are discussed in
the Einstein-Maxwell theory

\begin{equation}
S=\int\mathrm{d}^{4}x\sqrt{-g}\left[\frac{2}{\kappa^{2}}R-\frac{1}{2}g^{\mu\nu}g^{\alpha\beta}F_{\mu\nu}F_{\alpha\beta}\right],\label{eq:EH}\end{equation}
where $R$ is the Ricci scalar, $\kappa^{2}=32\pi G_{\mathrm{N}}$
and $F_{\mu\nu}$ denotes the $U(1)$ field strength tensor. Quantum
effects are calculated in the weak field expansion around the flat
Minkowski metric ($\eta_{\mu\nu}=(1,-1,-1,-1)$)

\begin{equation}
g_{\mu\nu}=\eta_{\mu\nu}+\kappa h_{\mu\nu}(x).\label{eq:lin1}\end{equation}
This is considered an exact relation, but the inverse of the metric
contains higher order terms, \begin{equation}
g^{\mu\nu}=\eta^{\mu\nu}-\kappa h^{\mu\nu}+\kappa^{2}h_{\alpha}^{\mu}h^{\nu\alpha}+\ldots\,,\label{eq:lin2}\end{equation}
in an effective treatment it can be truncated at the second order.
The photon propagator is defined in Landau gauge and the graviton
propagator in de Donder (harmonic) gauge, the condition is (with $h=h_{\alpha}^{\alpha}$)\begin{equation}
\partial^{\nu}h_{\mu\nu}-\frac{1}{2}\partial_{\mu}h=0.\label{eq:gauge}\end{equation}
Expanding the Lagrangian up to second order in the graviton field
the graviton propagator in $d$ dimensions is \begin{equation}
G_{\alpha\beta\gamma\delta}^{G}(k)=i\frac{\frac{1}{2}\eta_{\alpha\gamma}\eta_{\beta\delta}+\frac{1}{2}\eta_{\alpha\delta}\eta_{\beta\gamma}-\frac{1}{d-2}\eta_{\alpha\beta}\eta_{\gamma\delta}}{k^{2}-i\epsilon}.\label{eq:propg}\end{equation}
There are two relevant vertices involving two photons. The two photon-graviton
vertex is\begin{eqnarray}
V_{\gamma\gamma G}(k_{1\mu},k_{2\nu},\alpha,\beta) & \!=\! & -i\frac{\kappa}{2}\left[\eta_{\alpha\beta}\left(k_{1\nu}k_{2\mu}-\eta_{\mu\nu}(k_{1}k_{2})\right)+\right.\nonumber \\
 &  & \!\left.+Q_{\mu\nu,\alpha\beta}(k_{1}k_{2})+Q_{k_{1}k_{2},\alpha\beta}\eta_{\mu\nu}-Q_{\mu k_{2},\alpha\beta}k_{1\nu}-Q_{k_{1}\nu,\alpha\beta}k_{2\mu}\right],\label{eq:V3}\end{eqnarray}
and the two photon-two graviton vertex is even more complicated

\begin{eqnarray}
V_{\gamma\gamma GG}(k_{1\mu},k_{2\nu},\alpha,\beta,\gamma,\delta) & \!=\! & -i\frac{\kappa^{2}}{4}\left[P_{\alpha\beta\gamma\delta}\left(k_{1\nu}k_{2\mu}-\eta_{\mu\nu}(k_{1}k_{2})\right)+U_{\mu\nu,\alpha\beta,\gamma\delta}(k_{1}k_{2})+\phantom{\kappa^{2}}\right.\nonumber \\
 &  & \!+U_{k_{1}k_{2},\alpha\beta,\gamma\delta}\eta_{\mu\nu}-U_{\mu k_{2},\alpha\beta,\gamma\delta}k_{1\nu}-U_{k_{1}\nu,\alpha\beta,\gamma\delta}k_{2\mu}+\nonumber \\
 &  & \!+Q_{\mu\nu,\alpha\beta}Q_{\gamma\delta,k_{1}k_{2}}+Q_{\mu\nu,\gamma\delta}Q_{\alpha\beta,k_{1}k_{2}}\nonumber \\
 &  & \!\left.-Q_{k_{1}\nu,\alpha\beta}Q_{\mu k_{2},\gamma\delta}-Q_{\mu k_{2},\alpha\beta}Q_{k_{1}\nu,\gamma\delta}\phantom{\kappa^{2}}\right].\label{eq:V4}\end{eqnarray}
For the sake of simplicity we have defined \begin{equation}
U_{\mu\nu,\alpha\beta,\gamma\delta}=\eta_{\mu\alpha}P_{\nu\beta,\gamma\delta}+\eta_{\mu\beta}P_{\alpha\nu,\gamma\delta}+\eta_{\mu\gamma}P_{\alpha\beta,\nu\delta}+\eta_{\mu\delta}P_{\alpha\beta,\gamma\nu},\label{eq:UU}\end{equation}
\begin{equation}
P_{\alpha\beta,\mu\nu}=\eta_{\mu\alpha}\eta_{\nu\beta}+\eta_{\mu\beta}\eta_{\nu\alpha}-\eta_{\mu\nu}\eta_{\alpha\beta},\label{eq:PP}\end{equation}
and finally\begin{equation}
Q_{\alpha\beta,\mu\nu}=\eta_{\mu\alpha}\eta_{\nu\beta}+\eta_{\mu\beta}\eta_{\nu\alpha}.\label{eq:QQ}\end{equation}

There are two graphs contributing to the photon self energy (Fig.
1.) with two vertices \eqref{eq:V3} giving $\Pi^{(a)}$ and one 4-leg
vertex \eqref{eq:V4} providing $\Pi^{(b)}$. We calculated the finite
and divergent parts of the 2-point function with improved cutoff,
naive 4-dimensional momentum cutoff and dimensional regularization.
For comparison using the technique of dimensional regularization with
different assumptions about treating the number of dimensions $d$
in the propagator and vertices various quadratically divergent cutoff
results can be identified using \eqref{eq: quad}.

\begin{figure}
\begin{centering}
\includegraphics[height=3cm]{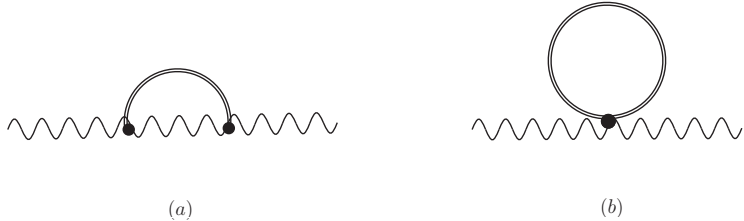}
\par\end{centering}

\centering{}\caption{Feynman graphs with graviton lines contributing to the photon two
point function.}

\end{figure}

The $\Lambda^{2}$ contributions of the two graphs with \textbf{improved
cutoff }(I) do not cancel each other \begin{eqnarray}
\Pi_{\mu\nu}^{\mathrm{I}(a)}(p) & = & \frac{i}{16\pi^{2}}\kappa^{2}\left(p^{2}\eta_{\mu\nu}-p_{\mu}p_{\nu}\right)\left(-2\Lambda^{2}-\frac{1}{6}p^{2}\left(\ln\left(\frac{\Lambda^{2}}{p^{2}}\right)+\frac{2}{3}\right)\right),\label{eq:impi1}\\
\Pi_{\mu\nu}^{\mathrm{I}(b)}(p) & = & \frac{i}{16\pi^{2}}\kappa^{2}\left(p^{2}\eta_{\mu\nu}-p_{\mu}p_{\nu}\right)\left(\phantom{2}\frac{3}{2}\Lambda^{2}\right).\label{eq:impi2}\end{eqnarray}

In the \textbf{naive cutoff} (N) calculation using \eqref{eq:negyed}
there is a cancellation of the $\Lambda^{2}$ terms, the finite term
differs from the previous one and it is impressing that the result
is transverse without any subtractions

\begin{eqnarray}
\Pi_{\mu\nu}^{\mathrm{N}(a)}(p) & = & \frac{i}{16\pi^{2}}\kappa^{2}\left(p^{2}\eta_{\mu\nu}-p_{\mu}p_{\nu}\right)\left(-\frac{3}{2}\Lambda^{2}-\frac{1}{6}p^{2}\ln\left(\frac{\Lambda^{2}}{p^{2}}\right)-\frac{7}{36}p^{2}\right),\label{eq:cupi1}\\
\Pi_{\mu\nu}^{\mathrm{N}(b)}(p) & = & \frac{i}{16\pi^{2}}\kappa^{2}\left(p^{2}\eta_{\mu\nu}-p_{\mu}p_{\nu}\right)\left(\phantom{-}\frac{3}{2}\Lambda^{2}\right).\label{eq:cupi2}\end{eqnarray}

In \textbf{dimensional regularization} (DR) the space-time dimension
is continued in all terms originating from the gauge and gravitational
part, too (e.g. $\eta_{\mu}^{\mu}=d=4-2\epsilon$). The result agrees
with \cite{rodigast} (without the finite terms which are first given
here) 

\begin{eqnarray}
\Pi_{\mu\nu}^{DR1(a)}(p) & = & \frac{i}{16\pi^{2}}\kappa^{2}\left(p^{2}\eta_{\mu\nu}-p_{\mu}p_{\nu}\right)\left(-\frac{1}{6}p^{2}\left(\frac{2}{\epsilon}+\ln\left(\frac{\mu^{2}}{p^{2}}\right)+\frac{1}{6}\right)\right),\label{eq:DRpi1}\\
\Pi_{\mu\nu}^{DR1(b)}(p) & = & 0,\label{eq:DRpi2}\end{eqnarray}
where we have omitted the constants $-\gamma_{E}+\ln4\pi$ beside
$2/\epsilon$. In what follows the various results using the technique
of dimensional regularization based on different assumptions are denoted
by the superscript $DR1,\, DR2,\, DR3$ and the corresponding cutoff
results by $\Lambda1,\,\Lambda2,\,\Lambda3$.

Now with the help of the equations \eqref{eq: quad}, \eqref{eq:log}
and \eqref{eq:finite} we can define a cutoff result based on the
dimensional regularization one. In principle the dimension has to
be modified in each terms where $d$ appears, also in the graviton
propagator \eqref{eq:propg}, though gravity is not a dynamical theory
in $d=2$. Each graph is quadratically divergent, even $1/(\epsilon-1)^{2}$
type of singularities appear in single graphs, but they cancel in
the sum of the graphs, like the $\frac{1}{\epsilon^{2}}$ terms in
usual gauge theories (e.g. in QCD) at two loops.\begin{equation}
\Pi_{\mu\nu}^{\mathrm{\Lambda1}}(p)=\frac{i}{16\pi^{2}}\kappa^{2}\left(p^{2}\eta_{\mu\nu}-p_{\mu}p_{\nu}\right)\left(-\frac{1}{4}\Lambda^{2}-\frac{1}{6}p^{2}\left(\ln\left(\frac{\Lambda^{2}}{p^{2}}\right)-\frac{5}{6}\right)\right)\label{eq:L1}\end{equation}
also quadratically divergent, but only the coefficient of the logarithmic
term agrees with other results.

The result of the improved momentum cutoff can be reproduced applying
dimensional regularization with care. The improved cutoff method works
in four physical dimensions and special rules have to be applied only
at the evaluation of the last tensor integrals. It is equivalent to
setting $d=4$ in the Einstein-Maxwell theory, e.g. both in the graviton
propagator and in the trace of the metric tensor. Dimensional regularization
is then applied at the last step evaluating the tensor and scalar
momentum integrals. We have found that $\Pi_{\mu\nu}^{\mathrm{DR2(b)}}=0$
and 

\begin{equation}
\Pi_{\mu\nu}^{DR2}(p)=\frac{i}{16\pi^{2}}\kappa^{2}\left(p^{2}\eta_{\mu\nu}-p_{\mu}p_{\nu}\right)\left(-\frac{1}{6}p^{2}\left(\frac{2}{\epsilon}+\ln\left(\frac{\mu^{2}}{p^{2}}\right)+\frac{5}{3}\right)\right).\label{eq:DR3}\end{equation}
 The corresponding cutoff result diverges quadratically and agrees
with the improved cutoff calculation

\begin{eqnarray}
\Pi_{\mu\nu}^{\mathrm{\Lambda2}(a)}(p) & = & \frac{i}{16\pi^{2}}\kappa^{2}\left(p^{2}\eta_{\mu\nu}-p_{\mu}p_{\nu}\right)\left(-2\Lambda^{2}-\frac{1}{6}\left(p^{2}\ln\left(\frac{\Lambda^{2}}{p^{2}}\right)+\frac{2}{3}\right)\right),\label{eq:DR3i1}\\
\Pi_{\mu\nu}^{\mathrm{\Lambda2}(b)}(p) & = & \frac{i}{16\pi^{2}}\kappa^{2}\left(p^{2}\eta_{\mu\nu}-p_{\mu}p_{\nu}\right)\left(\phantom{2}\frac{3}{2}\Lambda^{2}\right).\label{eq:DR3i2}\end{eqnarray}

To find connection with existing, partially controversial literature,
we have performed the calculation with weaker assumptions. First the
term in the graviton propagator is set $\frac{1}{d-2}=\frac{1}{2}$
as is usually done in earlier results e.g. \cite{don1,brazil}. The
divergent part of the dimensional regularization result agrees with
\cite{rodigast}. The contribution of the tadpole in Fig. 1b $\Pi^{\mathrm{DR3(b)}}$
vanishes, the sum is\begin{equation}
\Pi_{\mu\nu}^{DR3}(p)=\frac{i}{16\pi^{2}}\kappa^{2}\left(p^{2}\eta_{\mu\nu}-p_{\mu}p_{\nu}\right)\left(-\frac{1}{6}p^{2}\left(\frac{2}{\epsilon}+\ln\left(\frac{\mu^{2}}{p^{2}}\right)+\frac{1}{6}\right)\right).\label{eq:DR2}\end{equation}
We can identify a cutoff result, Fig. 1b gives $\Pi_{\mu\nu}^{\mathrm{\Lambda3(b)}}(p)\sim\frac{1}{2}\Lambda^{2}$,
the only quadratically divergent term,\begin{equation}
\Pi_{\mu\nu}^{\mathrm{\Lambda3}}(p)=\frac{i}{16\pi^{2}}\kappa^{2}\left(p^{2}\eta_{\mu\nu}-p_{\mu}p_{\nu}\right)\left(\frac{1}{2}\Lambda^{2}-\frac{1}{6}p^{2}\left(\ln\left(\frac{\Lambda^{2}}{p^{2}}\right)-\frac{5}{6}\right)\right).\label{eq:L2}\end{equation}
Notice that this result differs from \eqref{eq:L1} only in the coefficient
of the first term, the change comes from the different treatment of
the graviton propagator.

In principle a theory is completely defined via specifying the Lagrangian
and the method of calculation e.g. fixing the regularization and the
treatment of the divergent terms, though the physical quantities must
be independent of the details. It is remarkable that the transverse
structure of the photon propagator is not violated in any of the previous
regularizations and the logarithmic term is universal and agrees with
earlier results \cite{rodigast,deser}. The question is whether the
$\Lambda^{2}$ terms contribute to the running of the gauge coupling,
or not.

\section{Quadratic divergences and renormalization}

The 1-loop corrections to the 2-point function modify the bare Lagrangian,
the divergences have to be removed by the counterterms. Consider the
QED action with the convention \cite{don2}\begin{equation}
L_{0}=-\frac{1}{4e_{0}^{2}}F_{\mu\nu}F^{\mu\nu}+\bar{\Psi}iD_{\mu}\gamma^{\mu}\Psi,\qquad D_{\mu}=\partial_{\mu}+iA_{\mu}.\label{eq:Lqed}\end{equation}
The divergences calculated from the interaction \eqref{eq:EH} modify
the bare Lagrangian\begin{equation}
L=-\frac{1+a\kappa^{2}\Lambda^{2}}{4e_{0}^{2}}F_{\mu\nu}F^{\mu\nu}+a_{2}\ln\frac{\Lambda^{2}}{p^{2}}\left(D_{\mu}F^{\mu\nu}\right)^{2}+\left(\bar{\Psi}iD_{\mu}\gamma^{\mu}\Psi,\right),\label{eq:LqedR}\end{equation}
where $p^{2}$ is the Euclidean momentum at which the 2-point function
was calculated. 

The divergent terms have to be canceled by the counterterms. New dimension-six
term must be added to match the $p^{2}\ln\left(\frac{\Lambda^{2}}{p^{2}}\right)$
term, \begin{equation}
L_{\mathrm{\mathrm{ct}}}=\frac{\delta Z_{1}}{4e_{0}^{2}}F_{\mu\nu}F^{\mu\nu}+\delta Z_{2}\left(D_{\mu}F^{\mu\nu}\right)^{2}.\label{eq:ct}\end{equation}
In principle there are three possible dimension-six counterterms,
two of them linearly independent and it turns out that the second
term in \eqref{eq:ct} can cancel all divergences. The coefficient
of the first term in \eqref{eq:LqedR} cannot be understood as defining
a running coupling but it is compensated by a counterterm through
a renormalization condition. It can be fixed either by the Coulomb
potential or Thomson scattering at low energy identifying the usual
electric charge as\begin{equation}
\frac{e_{0}^{2}}{4\pi(1+a\kappa^{2}\Lambda^{2})}=\frac{e^{2}}{4\pi}\simeq\frac{1}{137}.\label{eq: alpha}\end{equation}
The quadratically divergent correction defines the relation between
the bare charge $e_{0}(\Lambda)$ in a theory with the physical cutoff
$\Lambda$ and the physical charge effective at low energies. After
fixing the parameters of the theory (e.g. by a measurement at low
energy) and using $e$ to calculate the predictions of the model the
cutoff dependence completely disappears from the physical charge \cite{don1,don2}.
The role of the quadratic correction is to define the relation\eqref{eq: alpha}
and to renormalize the bare coupling constant $e_{0}(\Lambda)$. (
and does not appear in the physical charge.)

The logarithmically divergent contribution on the other hand defines
the renormalization of the higher dimensional operator $\left(D_{\mu}F^{\mu\nu}\right)^{2}$
and again not the running of the gauge coupling. After renormalization
(at a point $p^{2}=\mu^{2}$) the logarithmic coefficient of the dim-6
term in\eqref{eq:LqedR} changes to $a_{2}\ln\frac{\Lambda^{2}}{p^{2}}-a_{2}\ln\frac{\Lambda^{2}}{\mu^{2}}=-a_{2}\ln\frac{p^{2}}{\mu^{2}}$
defining a would be running parameter. Furthermore note that\cite{rodigast,ellis}
this term can be removed by local field redefinition of $A_{\mu}$
up to higher dimensional operators\begin{equation}
A_{\mu}\rightarrow A_{\mu}-c\nabla_{\nu}F_{\mu}^{\nu},\label{eq:redef}\end{equation}
 where $\nabla_{\mu}$ is the gravitational covariant derivative,
as the new term is proportional to the tree level equation of motion
$D_{\mu}F^{\mu\nu}=0$. Generally, all photon propagator corrections
can be removed by appropriate field redefinition which are bilinear
in $A_{\mu}$ even if they contain arbitrary number of derivatives,
on-shell scattering processes are not influenced by the presence of
such effective terms \cite{ts} .

\section{Conclusions }

We have calculated the gravitational corrections to the simplest,
$U(1)$ gauge theory. This paper was motivated by the various, sometime
controversial results in the literature. Our method was capable of
identifying quadratically divergent contributions to the photon two
point function, thanks to the construction in a gauge invariant way.
To test our calculation we defined the cutoff dependence employing
\eqref{eq: quad},\eqref{eq:log} anddimensional regularization with
various assumptions about treating the number of dimensions $d$.
We observed that the 1-loop gravity corrections to the two point function
in all but one cases contain $\Lambda^{2}$ divergence with the exception
of the naive momentum cutoff which violates gauge symmetries usually.
Here all the corrections are transverse and the logarithmic term universally
agrees with the literature starting from Deser et al. \cite{deser}.

The quadratically divergent corrections to the photon self-energy
do not lead to the modification of the running of the gauge coupling.
Robinson and Wilczek claimed that the $-a\kappa^{2}\Lambda^{2}$ correction
could turn the beta function negative and make the Maxwell-Einstein
theory asymptotically free. This statement and the calculation was
criticized in the literature. We showed in this paper using explicit
cutoff calculation that $\Lambda^{2}$ corrections do appear in the
2-point function, but those will define the renormalization connection
between the cutoff dependent bare coupling and the physical coupling
\eqref{eq: alpha} and do not lead to a running coupling. Indeed the
$\Lambda^{2}$ correction absorbed into the physical charge and does
not appear in physical processes. Donoghue et al. argue in \cite{don1}
that a universal, i.e. process independent running coupling constant
cannot be defined in the effective theory of gravity independently
of the applied regularization. They demonstrate that because of the
crossing symmetry in theories (except the $\lambda\Phi^{4}$) even
the sign of the quadratic running is ambiguous and a running coupling
would be process dependent. Generally the logarithmically divergent
corrections could define the renormalization of higher dimensional
operators. It turns out that even these logarithmic correction can
be removed by appropriate field redefinitions and do not contribute
to on-shell scattering processes. We note that the authors in \cite{brazil}
showed using their 4-dimensional implicit regularization method that
the quadratic terms are coming from ambiguous surface terms and as
such are non-physical. Interestingly those surface terms vanish if
we evaluate them with our improved cutoff.

Finally we point out that we have found gravity corrections to the
two point function of the $U(1)$ gauge theory. Using a momentum cutoff
the quadratically divergent contributions define the renormalization
of the bare charge and thus using the physical charge the $\Lambda^{2}$
corrections do not appear in physical processes. On the other hand
logarithmic corrections are universal but merely define the renormalization
of a dimension six term in the Lagrangian, which term can be eliminated
by local field redefinition. We conclude that gravity corrections
do not lead to the modification of the usual running of gauge coupling
and cannot point towards asymptotic freedom in the case of electromagnetism.

\appendix

\section{The divergent integrals}

The loop integrals are calculated as follows. First the loop momentum
($k$) integral is Wick rotated (to $k_{E}$), with Feynman parameter(s)
the denominators are combined, then the order of Feynman parameter
and the momentum integrals are changed. After that the loop momentum
($k_{E}\rightarrow l_{E}$) is shifted to have a spherically symmetric
denominator.

In this appendix we present two divergent integrals calculated by
the new regularization. $\Delta$ can be any loop momentum independent
expression depending on the Feynman $x$ parameter, external momenta,
masses, e.g. $\Delta(x,q_{i},m).$ The integration is understood for
Euclidean momenta with absolute value below $\Lambda$.

The integral \eqref{eq:A1} is just given for comparison, it is calculated
with a simple momentum cutoff. In \eqref{eq:A2} with the standard
\eqref{eq:negyed} substitution one would get a constant $-\frac{3}{2}$
instead of $-1$ \cite{uj}.

\begin{eqnarray}
\int_{\Lambda\, reg}\frac{d^{4}l_{E}}{i(2\pi)^{4}}\frac{1}{\left(l_{E}^{2}+\Delta^{2}\right)^{2}} & \!\!=\!\! & \frac{1}{(4\pi)^{2}}\left(\ln\left(\frac{\Lambda^{2}+\Delta^{2}}{\Delta^{2}}\right)+\frac{\Delta^{2}}{\Lambda^{2}+\Delta^{2}}-1\right).\label{eq:A1}\\
\int_{\Lambda\, reg}\frac{d^{4}k}{i(2\pi)^{4}}\frac{l_{E\mu}l_{E\nu}}{\left(l_{E}^{2}+\Delta^{2}\right)^{3}} & \!\!=\!\! & \frac{1}{(4\pi)^{2}}\frac{g_{\mu\nu}}{4}\left(\ln\left(\frac{\Lambda^{2}+\Delta^{2}}{\Delta^{2}}\right)+\frac{\Delta^{2}}{\Lambda^{2}+\Delta^{2}}-1\right).\label{eq:A2}\end{eqnarray}

\end{document}